\documentclass[12pt,authoryear]{elsarticle}

\usepackage{amssymb}
\usepackage{amsmath}

\begin{document}

\begin{frontmatter}



\title{Wealth dynamics in a multi-aggregate \\closed monetary system}


\author[inst1]{Andrea Monaco}



\affiliation[inst1]{organization={University College Dublin, School of Mathematics and Stat.},
            addressline={Belfield}, 
            city={Dublin 4},
            country={Ireland}}

\author[inst2]{Matteo Ghio}
\author[inst1]{Adamaria Perrotta}


\affiliation[inst2]{organization={Bid Company},
            addressline={Viale Vincenzo Lancetti, 43}, 
            city={Milan},
            postcode={20158}, 
            country={Italy}}

\begin{abstract}
We examine the statistical properties of a closed monetary economy with multi-aggregates interactions. Building upon Yakovenko's single-agent monetary model \citep{Yako1}, we investigate the joint equilibrium distribution of aggregate size and wealth. By comparing theoretical and simulated data, we validate our findings and investigate the influence of both micro dynamics and macro characteristics of the system on the distribution. Additionally, we analyse the system's convergence towards equilibrium under various conditions. Our laboratory model may offer valuable insights into macroeconomic phenomena allowing to reproduce typical wealth distribution features observed in real economy.
\end{abstract}




\end{frontmatter}


\section{Introduction}
\label{sec:sample1}

The study of wealth dynamics has always held a central position in the field of economics \citep{sargan, pfeffer, kuhn}. This is because many questions posed by various economic theories, of significant social interest, are closely tied to wealth dynamics. Some examples of these questions, which carry substantial relevance, include the origins of social disparities, the factors influencing social mobility throughout history, and methods for creating a sustainable and more socially equitable economy. It is evident that addressing these questions requires a thorough exploration of wealth dynamics.

The prominence of wealth study within the scientific discourse among various economic theories is exemplified by the enduring popularity of economists who have focused on wealth analysis over time. Renowned works like \textit{An Inquiry into the Nature and Causes of the Wealth of Nations} \citep{smith}, \textit{The Capital} \citep{marx}, and more recently, Piketty's book on inequality titled \textit{The Economics of Inequality} \citep{piketty}, have all enjoyed widespread acclaim.

Furthermore, alongside the historical interest in studying wealth dynamics, there has been a renewed focus on this topic in recent years \citep{saez, benhabib, achdou}, since, we have access to historical time series data that tracks the evolution of wealth. Additionally, the analysis of these new wealth data-sets reveals a remarkably consistent pattern in wealth distribution, following an exponential trend \citep{hernandez}. This characteristic, observed in the past \citep{exp1, exp2, exp3}, has sparked increased interest in the subject today. The unexpected regularity in wealth distribution is particularly noteworthy, considering the numerous and complex factors suggested by economic theories that influence wealth dynamics \citep{macroIncome}.

Finally, in addition to the modeling and phenomenological interest in describing wealth distribution, a practical aspect has emerged concerning the global inflationary crisis. Indeed, central banks employ monetary policy to counter rising inflation \citep{Inflo1, Inflo2}. Operating through the banking system, this approach influences a vital aspect of wealth, commonly referred to as \textit{liquidity}. Therefore the effectiveness of central banks' monetary policies in addressing inflation relies on a thorough understanding of the mechanisms governing this wealth component (i.e., the \textit{liquidity}).

These reasons mentioned above for the growing interest in studying wealth dynamics have also led to the adoption of new approaches in researching this subject. Specifically, even in a field traditionally dominated by conventional macroeconomic models \citep{RevMicro1, RevMicro2, RevMicro3}, a micro-founded approach has started to gain traction. This approach recognizes the importance of individual-level dynamics in influencing macroeconomic outcomes.

Among these new approaches, a fundamental contribution has come from the application of methods typically employed in the so called exact sciences, such as physics. This approach, known as econophysics \citep{EcoPhysDef}, has emerged as a powerful framework for bridging the gap between micro and macro phenomena. By borrowing concepts and techniques from physics, econophysics seeks to uncover underlying patterns, dynamics, and principles governing economic systems.


Econophysics has significantly advanced our understanding of complex economic systems, providing insights into market dynamics, income distribution, and systemic risks \citep{Eco3Contrib1, Eco3Contrib2, Eco3Contrib3a, Eco3Contrib3b, Eco3Contrib3c}. By applying methods from statistical physics and complexity science, econophysics offers an interesting perspective on financial market patterns, asset price fluctuations, wealth inequality, and the interplay between individual behaviors and macroeconomic outcomes \citep{EcoRev1, EcoRev2}.

An example of applying physical sciences to the study of an economic phenomenon is the Yakovenko's work on wealth distribution \citep{Yako1}. In this pioneering study, Yakovenko addresses the issue of money distribution using a model where individual agents randomly exchange amounts of money, while considering an external constraint related to the isolation of the system (referred as \textit{microcanonical} ensemble in litterature \citep{MicroEnsemble}). 

With this simple model, we observe an equilibrium distribution that converges to an exponential shape, consistent with the predictions of the Boltzmann distribution within the microcanonical ensemble independently of the specific interaction assumed among individual agents. This result of Yakovenko found a partial experimental confirmation from works on the analysis of the income distribution \citep{EsDistrib1,EsDistrib2, EsDistrib3, EsDistrib4}.

Despite the importance of Yakovenko's work, it has taken time for the scientific community to fully embrace this approach. 
This delay can be attributed to the economics community's traditional preference for conventional methods when addressing extensively studied topics in the field. However, in recent years, the increased accessibility of data has expedited the transition to a new quantitative culture, enabling the exploration of innovative approaches to address the most relevant research topics in economics.

Our study aligns with the growing interest in new approaches to investigate macroeconomic phenomena. Specifically, we present an extended model based on Yakovenko's framework, incorporating the concept of agent aggregates within the microcanonical framework. 

Studying aggregates dynamics means studying a community's properties starting from basic hypotheses on the behaviour of their constituents (i.e. the single agents). Some examples in economics, of such duality, i.e. community and its constituents, are: countries and their economic sectors, economic sectors and their companies, companies and their stakeholder.

Our study focuses on examining the joint distribution of aggregate size and monetary wealth. In particular, analyzing the relationship between theoretical prediction and simulated data, we aim to uncover the influence of macro-level characteristics and micro-mechanisms on the wealth distribution. Furthermore, our study on the statistical equilibrium of economic systems addresses longstanding issues that have been central to economic debates, namely the dynamics of macro variables such as inflation \citep{Inflaz1, Inflaz2, Inflaz3}, as well as the effects of various monetary policies on wealth distribution \citep{ImpactMonePlicy1, ImpactMonePlicy2, ImpactMonePlicy3}.

From our perspective, the model presented here goes beyond being a mere monetary model. Its versatility and flexibility allow it to serve as a building block for models that encompass various components of agent-based systems. For instance, it can be seamlessly integrated into models investigating inflation dynamics or examining the impact of monetary policies on the broader economy. By incorporating wealth exchange as just one aspect, our model offers a comprehensive framework for exploring a wide range of macroeconomic phenomena.

\section{The aggregate model}

As stated above, our work builds upon the foundation established by Yakovenko \citep{Yako1}, enriching and transforming it with a new level of complexity introduced by the concept of aggregation. In this study, we analyze the statistical properties of a pure monetary closed economy, where numerous agents interact based on specific assumptions.
The pure monetary closed economy of Yakovenko consists of multiple agents, with each agent possessing an initial level of wealth that can be exchanged with other agents. The total amount of wealth remains constant. 

\begin{figure}
\includegraphics[scale=1.0]{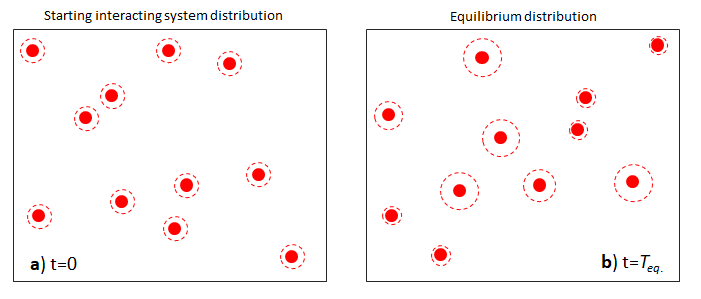}
\caption{
In figure a (left) and b (right), two different time configurations of the same single-agent monetary system are represented. On the left the starting configuration, on the right the equilibrium configuration. Each red dot represent a different agent while the size of the dotted circle indicates the wealth of the agent.}
\label{fig_yako_model}
\end{figure}

Within the Yakovenko model, the agents mutually interact randomly. Specifically, each agent, denoted as $i$, has a certain amount of money, $m_i$, which can be exchanged with other agents, denoted as $j$. The quantity of money exchanged, $\Delta m$, and the occurrence of the interaction between agents are both random variables. When money is exchanged, the impact on the wealth of agents $i$ and $j$ is represented as $[m_i, m_j] \rightarrow [m'_i, m'_j] = [m_i - \Delta m, m_j + \Delta m]$. It is important to note that the total amount of money remains conserved in each transaction: $m_i + m_j = m'_i + m'_j$. Figure \ref{fig_yako_model} represents a pure monetary economy, where each agent's wealth level is indicated by the size of the dotted red circle. The larger the circle, the higher the associated wealth level.

To enhance the pure monetary economy model proposed by Yakovenko and move towards a more comprehensive and realistic depiction of monetary dynamics, we introduce the concept of aggregate. An aggregate refers to a group of agents sharing similar characteristics, and each aggregate can interact with other aggregates according to specific rules.

In the aggregate model, every agent belongs to an aggregate, and the number of aggregates is significantly smaller than the number of agents. The initial size distribution of the aggregates, as well as the distribution of wealth, can follow different patterns.

Regarding the interactions within the model, we assume that the system includes two mechanisms:

\begin{itemize}

\item	{\textit{monetary wealth exchange}: each agent ($i$) can exchange a random quantity of money ($\Delta m$), sampled by a uniform distribution, with another agent ($j$, where $j \neq i$): $[m_i, m_j] \rightarrow [m'_i, m'_j ] = [m_i - \Delta m, m_j + \Delta m]$.} 
\item   {\textit{aggregate interaction}: each aggregate ($d_i$) can interact with another aggregate ($d_j$), where $j \neq i$, exchanging a random number of agents $\Delta n \in [0, n_{i}]$, where $n_i$ is the number of agents inside one aggregate. $[d_i, d_j] \rightarrow [d'_i, d'_j ]=[d_i - \Delta n, d_j + \Delta n]$}.

\end{itemize}

\begin{figure}
\includegraphics[scale=1.0]{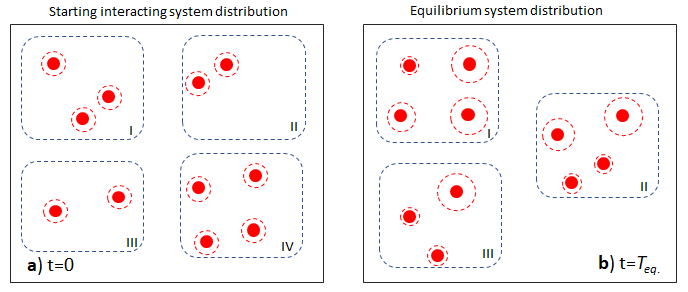}
\caption{
In figure a (left) and b (right), two different time configurations of the same aggregate monetary system are represented. On the left the starting configuration of the system with 4 aggregates, on the right the equilibrium configuration. Note that the content of each aggregate changes in terms of wealth and number of agents.}
\label{fig_agg_model}
\end{figure}

In Figure \ref{fig_agg_model}, we provide a graphical representation of the aggregate model. On the left side, we depict the initial configuration of agents grouped into four aggregates. On the right side, we observe the equilibrium distribution, where interactions, including those between aggregates, have resulted in a reduction to three aggregates. Furthermore, the composition of each aggregate has undergone changes in terms of both the number of agents and the overall wealth distribution.

Comparing Figure \ref{fig_agg_model} to Figure \ref{fig_yako_model}, we can observe two significant elements of novelty in the model. First, it introduces the concept of aggregate, representing groups of agents, which contributes to the notion of locality defined by the belonging to each aggregate. Second, the model incorporates interactions not only among agents but also between aggregates, expanding the scope of the analysis.

These two features are essential in bringing the model closer to its application and specialization in describing real-world phenomena. The new layer of interaction introduced with aggregates allows us to distinguish between intra-aggregate dynamics at a local scale, and extra-aggregate dynamics that involve macro-entities such as the aggregates themselves.

There are numerous examples that illustrate how the level of complexity in the proposed model aligns with the description of real-world phenomena \citep{ExComplex1a, ExComplex1b, ExComplex1c}. Consider, for instance, the dynamics of macroeconomic variables, which are often analyzed at both the individual community level (e.g., individual states or companies) and the larger community level (aggregates of states or industry sectors) \citep{ExComplex2a, ExComplex2b, ExComplex2c}. In these cases, the differences in behavior across different levels often become the main focus of investigation.

On the other hand, the key innovation and strength of our work are in integrating macroscopic observations into a micro-founded model through the introduction of aggregate dynamics.

In line with this approach, we have chosen to fully embrace the \textit{microcanonical} constraint \citep{MicroEnsemble}. This allows us to thoroughly analyze the relationships among variables by incorporating specific constraints from the microcanonical approach into the system.

\newpage

\section{The equilibrium distribution}

In the following, we will explore the equilibrium distribution of the system in terms of both monetary wealth and the size of individual aggregates. Specifically, we will examine the equilibrium distribution of the system under two regimes, characterized by a high number and a limited number of aggregates, respectively.

\subsection{The limit for a large number of aggregates}
\label{sec:distrib1}

Hereafter, we derive the joint equilibrium probability distribution $p(m,d)$ of the size of aggregates ($d$) and monetary wealth ($m$), in the model introduced above.
Let us consider a system comprising a constant number of $N_{a}$ aggregates, of $D$ agents, and a total monetary mass indicated by $M$. The size of aggregates ($d$) and monetary wealth ($m$) are distributed among the aggregates. Denoting with $n_{m,d}$ the number of aggregates with wealth $m$ and size $d$, the following conservation relationships hold for: the number of aggregates, the total wealth, and the number of individuals:

\begin{equation}
\sum_{m,d} n_{m,d} = N_a, \quad \sum_{m,d} n_{m,d} \cdot d = D, \quad  \sum_{m,d} n_{m,d} \cdot m = M.
\label{eq:1}
\end{equation}

Subject to the appropriate normalization, the probability of realizing a specific configuration (i.e., partition) of the system in terms of monetary mass and number of agents within each individual aggregate is given by:

\begin{equation}
W(n)=\frac{N_a!}{\prod_{m,d} n_{m,d}!}\prod_{m,d}\binom{m+d-1}{d-1} ^{n_{m,d}},
\end{equation}
where $W(n)$ represents the number of ways to arrange $N_a$ different aggregates in all possible accessible locations, each characterized by a specific value of wealth $m$ and size $d$. The binomial coefficient indicates the multiplicity, i.e., the way of distributing $m$ coins (units of monetary wealth) among $d$ individuals within a single aggregate in different ways. Each of these coefficients is then raised to the power of $n_{m,d}$. To derive the equilibrium distribution, we employ the standard procedure of statistical mechanics. This involves maximizing the system's entropy, and as a result, identifying the distribution associated with this maximum entropy as the equilibrium distribution \citep{MicroEnsemble}. To this scope we take the logarithm of $W$ and utilizing Stirling's formula (i.e., $ln(n!) \approx n \cdot ln(n)-n$) \citep{Stirling} for large values of $n$, obtaining:

\begin{equation*}
\ln(W) = N_a  \ln(N_a)-N_a-\sum_{m,d}n_{m,d} \ln(n_{m,d})+\sum_{m,d}n_{m,d}+n_{m,d} \sum_{m,d}\ln\binom{m+d-1}{d-1}.
\end{equation*}
By exploiting the conservation of aggregates and differentiating both sides respect to $n_{m,d}$, along with the three conservation relationships (see eq.\ref{eq:1}), we have:

\begin{eqnarray*}
d\ln(W) = -\sum_{m,d}(\ln(n_{m,d}) + 1) dn_{m,d} + \sum_{m,d}\ln\binom{m+d-1}{d-1}dn_{m,d},\\
\sum_{m,d} dn_{m,d} = 0, \:\:\:\:\:\: \sum_{m,d} dn_{m,d} \cdot d = 0, \:\:\:\:\:\: \sum_{m,d} dn_{m,d} \cdot m = 0. 
\end{eqnarray*}

Using the method of Lagrange multipliers \citep{LagrangeMulty}, multiplying the conservation relationships of the number of aggregates, wealth, and individuals by $\alpha$, $\beta$ and $\gamma$, respectively, and then summing them with the existing equations $d\ln(W)$, we obtain:

\begin{eqnarray*}
\sum_{m,d}\left\{(\ln(n_{m,d}) + 1) + \ln\binom{m+d-1}{d-1} - \alpha - \beta m - \gamma d\right\}dn_{m,d} = 0.
\end{eqnarray*}

Hence, the equilibrium distribution is given by:

\begin{eqnarray}
p(m,d) = C\binom{m+d-1}{d-1}e^{-\beta m}e^{-\alpha d},
\label{eq:2}
\end{eqnarray}
where $C$ is the normalization constant.

We observe that, similar to the case without aggregates (i.e. the Yakovenko model), the distribution in this scenario is also exponential with respect to the two new variables: wealth ($m$) and the size of individual aggregates ($d$). The binomial coefficient that precedes the two exponential in Eq. \ref{eq:2} accounts for all possible distributions of wealth and the number of agents within each aggregate.
As for the significance of the Lagrange constants $\alpha$ and $\beta$, they are associated with macroscopic quantities of the system, such as the total wealth of the system ($M$) and the total number of agents ($D$). Computing the average wealth of the system, the average size of the aggregate, and considering the normalization condition of the distribution, we obtain the three following expressions for $C$, $\alpha$, and $\beta$ (see the Appendix for details):

\begin{eqnarray*}
C=\frac{N_a}{D+M}, 	\quad \alpha = -ln\left(\frac{D - N_a}{D + M}\right), \quad  \beta = -ln\left(\frac{M}{D+M}\right).
\end{eqnarray*}

Therefore the parameters of $p(m,d)$ are strictly dependent from the macroscopic properties of the system: $N_a$ (the total number of aggregates), $D$ (the total size of the system) and $M$ (the total wealth of the system).

\subsection{A limited number of aggregates}
\label{sec:theory_1}

Below, we propose an alternative approach to derive the equilibrium distribution of aggregates in terms of size and wealth. This approach enables us to explore the region where the value of $N_a$ is not significantly large. 

Indeed, in Section \ref{sec:distrib1} the Boltzmann distribution was derived from statistical mechanics and it is applicable to systems comprising a substantial number of particles (i.e. agents), a condition appropriate to a context that assumes an infinitely large number of particles (i.e., the thermodynamic limit). To be consistent with this hypothesis we assume that the system possesses an infinite volume, or significantly exceeds the characteristic length scales associated with the interactions among particles.

To create a more realistic model, we need to extract the equilibrium distribution by easing the constraint on the number of particles imposed by the statistical approach. Our objective now is to identify the equilibrium distribution of the system, considering a finite number of aggregates and agents. It is important to acknowledge that wealth exchange may not be entirely discrete. To accomplish this, we will utilize the properties associated with the Dirichlet function \citep{Dirichelet}.

Let us consider a system with a fixed number $D$ of agents, where the total wealth remains constant at 1. By denoting with $x_i$ the wealth of the $i$-th agent, we can express the probability distribution of the system's potential configurations (i.e. $f$), while adhering to the constraint of wealth conservation, using the following notation:

\begin{eqnarray}
f(x_1,..,x_i,..,x_D) = \frac{1}{B}\mathbf{1}_{\{ \sum_{i=1}^D x_i = 1  \}},
\label{eq:3}
\end{eqnarray}
with $B$ as the normalization constant. This formula can be seen as a particular case of the Dirichlet distribution function \citep{Dirichelet}:

\begin{eqnarray*}
f(x_1,..,x_i,..,x_D, \alpha_1,..,\alpha_i, .., \alpha_D) = \frac{1}{B(\alpha)}
\prod_{i=1}^D x_i^{\alpha_i - 1}\mathbf{1}_{\{\sum_{i=1}^D x_i = 1\}},
\end{eqnarray*}
where $B(\alpha)$ is given by:

\begin{eqnarray*}
B(\alpha) = \frac{\prod_{i=1}^D\Gamma(\alpha_i)}{\Gamma\left(\sum_{i=1}^D\alpha_i\right)}.
\end{eqnarray*}

In particular, if we set all values of $\alpha_i$ equal to 1 ($\alpha_i = 1$), effectively identifying $\alpha_i$ with the size of the $i$-th individual agent, we obtain eq.\ref{eq:3}. To determine the distribution function of the system with a fixed size of the individual aggregate, $d$, we can now leverage the aggregation property of the Dirichlet distribution \citep{Dirichelet}. In particular, if $(x_1,..,x_i,..,x_D )$ follows a Dirichlet distribution (i.e. $X  \sim Dir (\mathbf{\alpha}$), one can write:

\begin{eqnarray*}
(x_1,..,x_i+x_j,..,x_D )     \sim    Dir (\alpha_1,..,\alpha_i+\alpha_j,..,\alpha_D ).
\end{eqnarray*}

In other words, if we replace $x_i$ and $x_j$ with their sum (i.e. $x_i + x_j$), then $\alpha_i$ and $\alpha_j$ should also be replaced with their sum (i.e. $\alpha_i + \alpha_j$). This allows us to obtain the probability distribution that $d$ individuals form an aggregate with a relative wealth of $x = m/M$:

\begin{eqnarray*}
p(x)=\frac{\Gamma(D)}{\Gamma(d)\Gamma(D-d)} \left(x\right)^{d-1} \left(1-x\right)^{D-d-1} 1_{[0,1]}. 
\end{eqnarray*}

Or, by a change of variables:

\begin{eqnarray}
p(m)=\frac{\Gamma(D)}{\Gamma(d)\Gamma(D-d)} \left(\frac{m}{M}\right)^{d-1} \left(1-\frac{m}{M}\right)^{D-d-1} \frac{1}{M} 1_{[0,M]}. 
\label{eq_pm}
\end{eqnarray}
Now we consider the presence of $N_a$ aggregates in the system, interacting by exchanging $D$ individuals, and we assume that all the possible configurations are equally probable. The probability that a particular aggregate has a size of $d$ is given by:

\begin{eqnarray}
p(d)=  \frac{ \binom{D-d+N_a-1}{N_a-1}}{\binom{D+N_a-1}{N_a-1}}.
\label{eq_pd}
\end{eqnarray}

In the expression of $p(d)$, the denominator is determined by the number of ways we can arrange $D$ \textit{identical} individuals among $N_a - 1$ distinct aggregates. The numerator, on the other hand, is derived from the number of ways we can arrange the remaining $D-d$ individuals among the remaining $N_a - 1$ aggregates, once $d$ elements have already been placed in the specific aggregate under consideration. At this point, we can derive the joint distribution of wealth $m$ and size $d$ for the individual aggregate, as the process of exchanging the number of agents between aggregates is independent of the exchange of wealth. We get:

\begin{eqnarray}
p(m,d)= \frac{ \binom{D-d+N_a-1}{N_a-1}}{\binom{D+N_a-1}{N_a - 1}}\frac{\Gamma(D)}{\Gamma(d)\Gamma(D-d)} \left(\frac{m}{M}\right)^{d-1} \left(1-\frac{m}{M}\right)^{D-d-1} \frac{1}{M} 1_{[0,M]}. 
\end{eqnarray}

Similarly to the case of a large number of aggregates, in this scenario, the joint distribution $p(m,d)$ depends on the macroscopic characteristics of the system, such as $M$, $D$ and $N_a$.


\section{Simulating the system dynamics}

In the upcoming paragraphs, we will explore the outcomes obtained from numerical simulations of the system's dynamics. These simulations have successfully validated the equilibrium distribution derived in Section \ref{sec:theory_1}. Moreover, the simulations have enabled us to study the convergence of the system towards equilibrium over time.

\begin{figure}
\includegraphics[scale=0.55]{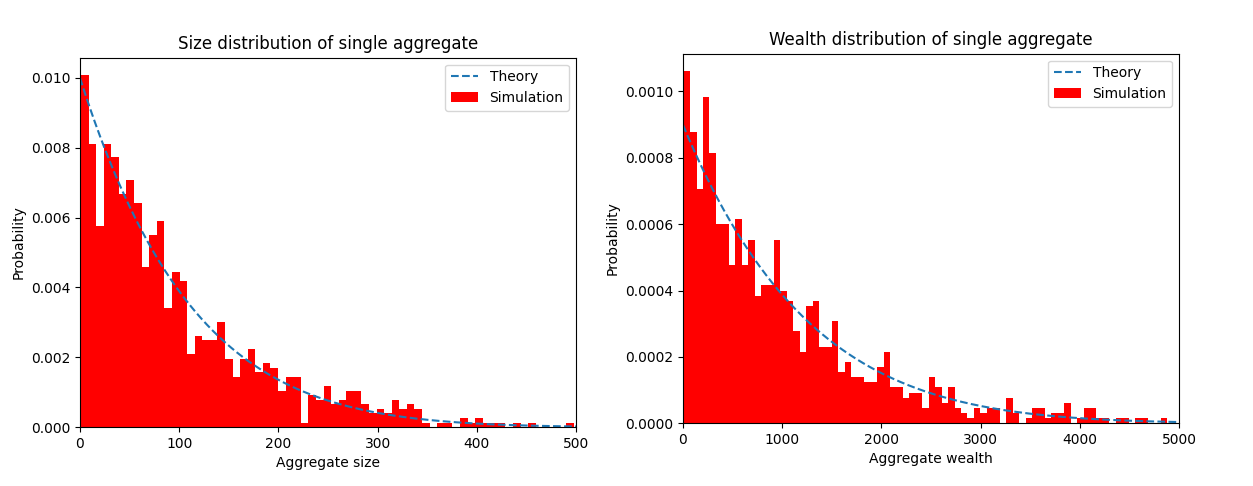}
\caption{
The Figures represent the size distribution (left) and wealth distribution (right) of the single aggregate. In both cases, the solid dotted line represents the prediction from the theoretical model (see eq. \ref{eq_pm}, \ref{eq_pd}).
}
\label{fig_distribution}
\end{figure}

\subsection{The system at equilibrium}

To gain insights into the properties of the model and validate the equilibrium distribution obtained in the previous sections, we conducted a simulation of the model's dynamics. In particular we simulated an initial configuration consisting of $1{,}000$ aggregates, with $100$ agents per aggregate, for a total of $100{,}000$ agents. Each aggregate was assigned a wealth of $10{,}000$, equivalent to $100$ units of wealth per individual agent.
The simulation adheres to the microcanonical constraint, where the system is considered isolated from the external environment and the total wealth of the system is conserved. 

The simulation progresses over time through a series of sequential operations. Firstly, a pair of interacting agents is randomly selected for wealth exchange. The amount of wealth transferred from one agent to another is determined randomly from a uniform distribution. Additionally, both interacting aggregates and the number of agents migrating between aggregates are determined through random drawing from uniform distributions.

In the course of the simulation, the wealth of each agent is updated, followed by the update of aggregate composition and, if needed, the adjustment of the number of aggregates. The number of aggregates can vary between one and the maximum number of agents in the system.

\begin{figure}
\centering
\includegraphics[scale=0.55]{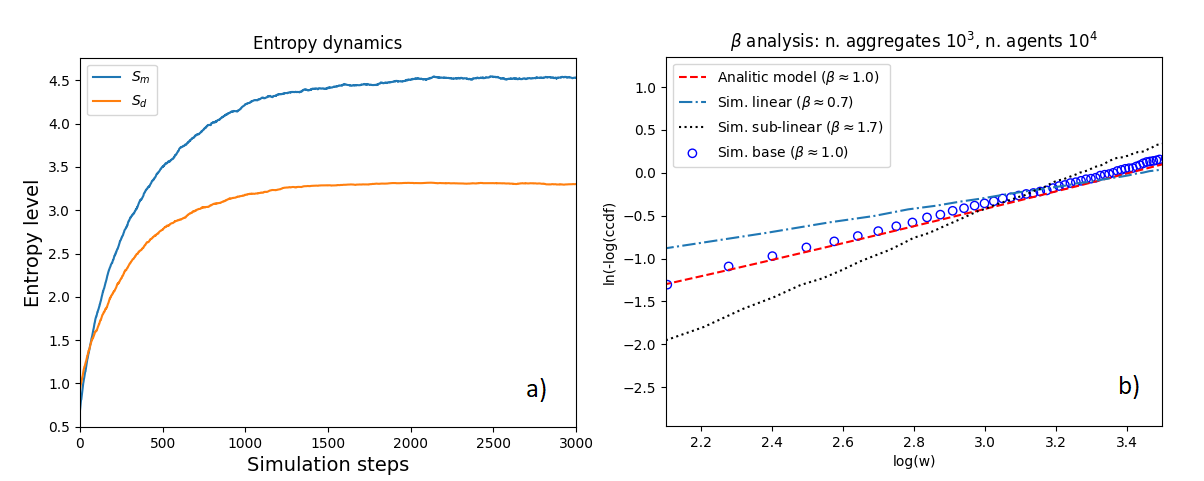}
\caption{
Figure a (left) depicts the convergence to the equilibrium of the size ($S_d$) and monetary ($S_m$) entropy as a function of the simulation step. In the Figure b (right) the dashed line represents the $\ln(-\log(\text{ccdf}))$ predicted by the theoretical model according the base interaction mechanism, while the open symbols depict the results of the simulation. The dot-dash line in Figure b is associated with the linear hypothesis on the exchange of agents among aggregates while the dotted line is the result of the sub-linear mechanism of exchange of agents.   
}
\label{fig_beta}
\end{figure}

We set to $4\times10^4$ the minimum number of simulation steps required to reach a state of equilibrium. This time attaining equilibrium is consistent with our observations in the subsequent analysis of the variation in the monetary and size entropy of the system. Monitoring entropy enables us to assess the system's proximity to equilibrium. In particular, from the data of Figure \ref{fig_beta} a (left), we deduce that the system reaches equilibrium after only 1{,}500 time steps of simulation, i.e. well before our simulation horizon of $4\times10^4$.

Once reached the last simulation time, we sample the sizes of the aggregates to reconstruct their equilibrium size distribution. Similarly, we sample the wealth quantities of each aggregate to reconstruct their wealth distribution. Both simulated distributions, namely the wealth distribution and the size distribution, are then compared with the theoretical predictions (see Figure \ref{fig_distribution}).
Therefore the comparison between the simulated joint distribution of aggregate size and monetary content provide a validation for the derived analytical equilibrium distribution. Moreover to test the reliability of the simulation, we conducted a comparison with Yakovenko's model in the case of a single aggregate, and our findings align with the results presented in Yakovenko's paper \citep{Yako1}.

Figure \ref{fig_distribution} provide insights into the distribution patterns within the model. Both the distributions of aggregate sizes and the distribution of monetary content within each aggregate exhibit an exponential pattern. This finding indicates that the introduction of aggregates as an additional layer of complexity does not deviate from the characteristic exponential shape observed in a pure monetary system.

Furthermore, to assess the robustness of the exponential shape, we consider the quantity $\log(-\log(\text{CCDF}))$, where $\text{CCDF} = 1 - \text{CDF}$ with $\text{CDF}$ denoting the cumulative distribution function. As showed by \citep{hernandez}, this quantity highlights the magnitude of the exponent $\beta$ in a stretched exponential distribution (i.e., $p(x) \propto x^{\beta - 1}e^{-(\frac{x}{x_0})^{\beta}}$).

Indeed, plotting $\log(-\log(\text{CCDF}))$ against $\log(x)$ reveals the magnitude of the $\beta$ parameter, resulting in the slope of the linear dependence between the two variables (i.e., $\log(-\log(\text{CCDF}))$ and $\log(x)$):

\begin{eqnarray*}
\text{CCDF} = 1- \text{CDF} = 1 - \int_0^{\infty}p(x)dx,\\
p(x) \propto x^{\beta - 1}e^{-(\frac{x}{x_0})^{\beta}}\rightarrow \text{CCDF} \propto   e^{-(\frac{x}{x_0})^{\beta}},\\
\\
\log(-\log(\text{CCDF})) \propto \beta \log (x) - \beta \log (x_0).
\end{eqnarray*}

Therefore, based on the previous relation, the right panel of Figure \ref{fig_beta} allows us to focus on the behaviour of the tail of the distribution, representing the strength of the $\beta$ parameter. In particular, the panel highlights the high level of agreement between the theoretical (dashed line) and simulated (open circle symbols) wealth distributions. Moreover, in the same panel, we present data corresponding to a simulation with two different micro-mechanisms of agent exchange, in addition to the base mechanism. These simulations are denoted as linear and sub-linear to distinguish them from the base mechanism.

In the base mechanism of agent exchange between aggregates, the quantity of agents exchanged, although random, has a constant expected value $\mathbf{E}[\Delta n] = \frac{1}{2} \hat{n}$ where $\hat{n}=100$ represents the number of agents per aggregate at the beginning of the simulation. This base setup results in a simulated value of $\beta \approx 1$ (dashed line in Figure \ref{fig_beta}).
Conversely, in the linear mechanism, the expected value of exchanged agents is not constant but equal to $\mathbf{E}[\Delta n]= \frac{1}{2} \hat{n}(t_i)$, with $\hat{n}(t_i)$ being the available number of agents of the aggregate at time $t_i$. The linear mechanism of agent exchange predicts a $\beta$ value equal to $0.7$ (dot-dashed line in Figure \ref{fig_beta}). In the case of the sub-linear mechanism, we have $\mathbf{E}[\Delta n]=\frac{1}{2}\hat{n}^{\gamma}(t_i)$, with $\gamma=0.9$, corresponding to a $\beta=1.7$ (dotted line in Figure \ref{fig_beta})

Therefore, Figure \ref{fig_beta} b (right), allows us to visualize how a simple change in the micro-mechanism of exchanging the number of agents between aggregates can alter the steepness of the distribution (i.e., $\beta$), transitioning from 1.7 to 0.7.

The influence of the agent exchange micro-mechanism on wealth distribution provides an illustration of the versatility of the proposed model in capturing diverse features of wealth distribution. Existing literature has demonstrated that the beta level of the distribution can vary over time and with different wealth indicators \citep{hernandez, exp1, exp2, exp3}. In this context, the model introduced here offers substantial flexibility in shaping interaction mechanisms between aggregates. These mechanisms can be customized based on the size or wealth of each aggregate. Additionally, interaction mechanisms between aggregates may differ, depending on whether they impact the exchange of agents or the quantity of wealth exchanged.

\subsection{Convergence to equilibrium of the system}

After examining the equilibrium distribution of the system, we proceeded to investigate its behavior outside of equilibrium. Our focus was on identifying the key variables that play a crucial role in driving the system towards equilibrium. This line of analysis is of significant interest due to its practical implications. Indeed, understanding the effectiveness and consequences of monetary policies implemented by Central Banks, particularly in relation to their impact on price distribution and overall inflation levels in the economy, remains an ongoing issue \citep{mpIneq, mpInfl}. In this context, it is essential, even with a first-principles model like the one proposed here, to discern the constraints that govern the dynamics of the money distribution when intentionally perturbing the system away from equilibrium.

To investigate these system features, we employed the concept of entropy \citep{DefEntropy}, focusing specifically on two variables that capture the dynamics of the system: wealth ($S_m$) and aggregate size ($S_d$). The following formulas were used to calculate these quantities:

\begin{eqnarray*}
S_m = -\sum_{i=1}^{N_m}p(m_i)\log(p(m_i)), \quad S_d = -\sum_{i=1}^{N_d}p(d_i)\log(p(d_i)).
\end{eqnarray*}
Here, $p(m_i)$ and $p(d_i)$ represent the probabilities of the $i$-th aggregate having a monetary quantity of $m_i$ or a size of $d_i$, respectively. $N_m$ and $N_d$ denote the number of states for wealth and aggregate size in the system. To discretize the continuous variable $m_i$ and compute $p(m_i)$, we sampled the distribution of wealth at each time step of the simulation and grouped the values into a histogram with fixed bin sizes. We then counted the number of occurrences of $m_i$ within each interval. However, discretization was not required for the computation of $p(d_i)$ since aggregate size is naturally discrete as it is determined by the number of agents within each aggregate.

\begin{figure}
\includegraphics[scale=0.55]{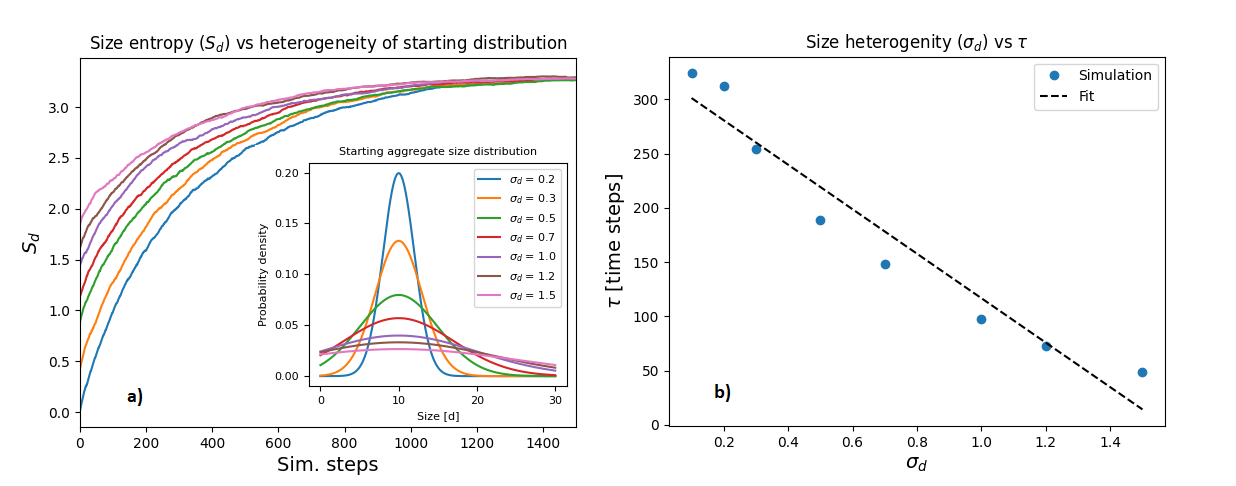}
\caption{
The Figure a (left) shows the system's entropy dynamics at increasing level of heterogeneity, here $\sigma_d$ indicates the standard deviation of the initial size distribution of the system. The insight into the left figure represent the size distribution at increasing levels of $\sigma_d$. The Figure b (right) shows the associate characteristic time convergence.} 
\label{fig_offeq_size}
\end{figure}

\begin{figure}
\includegraphics[scale=0.55]{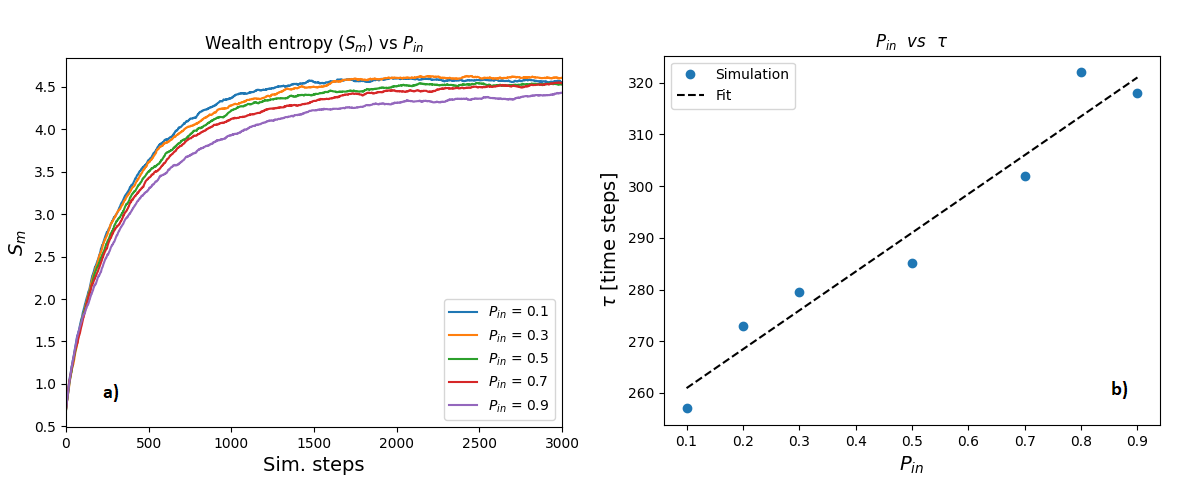}
\caption{
The figure a (left) shows the entropy dynamics at increasing levels of extra-aggregate interaction probability (i.e., $P_{in}$), while the figure b (right) presents the corresponding characteristic time convergence.}
\label{fig_offeq_prob}
\end{figure}

In Figures \ref{fig_beta}, we illustrate the convergence behavior of relative entropy for both aggregate size and wealth. The graph reveals a consistent trend towards a stable level. This analysis provides valuable insights, indicating that with a initial system configuration of approximately $1{,}000$ units of wealth per agent and aggregates containing $1{,}000$ agents, convergence to equilibrium is achieved within $1{,}400$ simulation steps. Furthermore, we observe that both quantities, $S_d$ (aggregate size entropy) and $S_m$ (wealth entropy), are influenced by the total wealth of the system ($M$) and the total number of agents ($D$). These findings align with the patterns observed in the purely monetary model.

We next studied the impact of system heterogeneity in terms of aggregate size on the convergence of the system to equilibrium. Specifically, we simulated systems with different levels of heterogeneity in their initial configurations by sampling the size of each aggregate from a normal distribution with increasing levels of standard deviation ($\sigma_d$), see Figure \ref{fig_offeq_size}.

Examining entropy dynamics, we note that the equilibrium entropy size of aggregates converges to a consistent level, regardless of the initial size heterogeneity within the system (see Figure \ref{fig_offeq_size} a). However, a higher initial size heterogeneity corresponds to an elevated initial entropy level. Consequently, this results in a reduced gap to the equilibrium entropy level and a quicker convergence time (see Figure \ref{fig_offeq_size} b).

Another aspect we investigated in terms of its impact on the convergence of the system to equilibrium is the probability of agents interacting exclusively within their own aggregates ($P_{in}$), without exchanging wealth with agents from different aggregates. To implement this mechanism, we introduced the possibility to control the probability of interaction between a given agent and an agent outside its own aggregate.

During the simulation, we categorized the aggregates into two groups: the aggregate containing the $i$-th agent, and all other aggregates excluding the $i$-th one. When selecting the $j$-th agent for interaction, we first determined the category of the $j$-th agent's aggregate based on a predefined probability. Subsequently, we randomly selected an agent from within that category.

The results show that the probability of interaction ($P_{in}$) between one agent and the other agents within its aggregate has a significant impact on the rate of convergence to equilibrium. Increasing the strength of $P_{in}$ results in a longer convergence time, indicating that reaching equilibrium takes more time when agents only interact with their immediate neighbours within the same aggregate.

To conduct a more precise quantitative analysis of the system's convergence time to equilibrium, we investigated the dynamics of simulated entropy levels using the following functional form: 

\begin{eqnarray*}
S_{m,d} = a \left(1 - e^{(-t/\tau)^{\xi}}\right).
\end{eqnarray*}

Here, $a$, $\xi$, and $\tau$ represent parameters derived from fitting each simulated entropy curve. The resulting values of $\tau$ obtained from the fitting procedure are presented in Figures \ref{fig_offeq_size} b and \ref{fig_offeq_prob} b.

From these figures, it is evident that both system's features, namely the heterogeneity and the infra-aggregate interaction probability, significantly influence the typical convergence times towards equilibrium. The observed strong linear relationship between $\tau$ and $\sigma_d$ as well as $P_{in}$ highlights the significant role these two parameters play in driving the wealth redistribution dynamics among individual agents.
Additionally, this relationship underscores their importance in the aggregation dynamics among different agent communities.

In other words, when devising a strategy for wealth redistribution within the system, it is essential to minimize barriers to interaction among agents, while also considering the construction of aggregates with different sizes, as this promotes a quicker approach to the equilibrium distribution of aggregate sizes.


\section{Conclusions}

In conclusion, we examined the statistical properties of a pure monetary closed economy, based on single agent and multi-aggregate dynamics. We derived the equilibrium joint distribution of aggregate size and wealth and we compared it with simulated data to validate our findings. Our analysis confirms the impact of macro-level characteristics and micro-mechanisms on the main features of wealth distribution as observed in real economies.

The outcomes of our analysis further expand upon and generalize the prior discoveries made by Yakovenko. Our results emphasize the enduring presence of an exponential distribution, even when complexity is introduced through aggregates. This underscores the significance of our study in comprehending real-world wealth distribution phenomena where this type of monetary distribution is observed \citep{ExComplex1a, ExComplex1b, ExComplex1c, hernandez}.

Furthermore, we investigated the convergence dynamics towards the equilibrium by studying the monetary and size entropy of the system. We measured the characteristic convergence time ($\tau$) under various conditions, including variations in the initial heterogeneity levels of each aggregate.

With this study we propose a new framework that combines aggregate and single-agent interactions, offering valuable insights into realistic economic systems. Indeed our model stands out for its realism and versatility compared to other approaches, extending beyond the mere ability to replicate the exponential pattern of wealth distribution.

Introducing the concept of aggregate, the model enables the specification of interaction mechanisms between aggregates and agents, allowing to link them to specific aggregate characteristics. This feature significantly enhances the model's ability to describe complex dynamics of wealth redistribution.

Analyzing the equilibrium distribution and entropy dynamics, we highlighted the roles of micro mechanisms in driving wealth distribution towards equilibrium. The research point out the limitations of economic systems in wealth redistribution and emphasizes the impact of different interaction mechanisms on equilibrium wealth levels.

Moreover, our exploration of different interaction mechanisms offers valuable insights for devising more effective monetary policies that consider the inherent tendency of wealth to exponentially concentrate.

Finally, we believe that in addition to advancing our understanding of monetary systems, the model presented here can serve as a foundational element for constructing models that incorporate various components of agent-based systems. This enables the exploration of more intricate economic phenomena.

\newpage


\section{Appendix}

\label{sec:sample:appendix}


In this section, we present the process of deriving the expression for the parameters of the probability distribution function. We will start by considering the distribution function:

\begin{eqnarray*}
p(m,d) = C\binom{m+d-1}{d-1}e^{-\beta m}e^{-\alpha (d-1)}, \sum_{m=0}^{\infty}\sum_{d=1}^{\infty}C\binom{m+d-1}{d-1}e^{-\beta m}e^{-\alpha (d-1)} = 1.
\end{eqnarray*}

Let’s compute the marginal probability:

\begin{eqnarray*}
p(d) = \sum_{m=0}^{\infty}C\binom{m+d-1}{d-1}e^{-\beta m}e^{-\alpha (d-1)},
\end{eqnarray*}

with $e^{-\beta}< 1$  we have:

\begin{eqnarray*}
p(d) = C e^{\alpha}\frac{e^{-\alpha d}}{(1 - e^{-\beta})^d},
\end{eqnarray*}

on the opposite case $e^{-\alpha}< 1$  we have:

\begin{eqnarray*}
p(m) = C e^{\beta}\frac{e^{-\beta (m + 1)}}{(1 - e^{-\alpha})^{(m+1)}}.
\end{eqnarray*}

Returning to the normalization of the probability:

\begin{eqnarray*}
C \sum_{d=1}^{\infty}e^{\alpha}\frac{e^{-\alpha d}}{(1 - e^{-\beta})^d} = 1, 
\end{eqnarray*}
needs
\begin{eqnarray*}
0 <\frac{e^{-\alpha}}{1 - e^{-\beta}} < 1,
\end{eqnarray*}
therefore we obtain:
\begin{eqnarray*}
Ce^{-\alpha}\frac{e^{-\alpha}}{1 - e^{-\beta}}\frac{1}{1 - \frac{e^{-\alpha}}{1 - e^{-\beta}}} = 1, \quad C\frac{1}{1 - e^{-\alpha} - e^{-\beta}} = 1.
\end{eqnarray*}

For the average size, the following condition must hold:

\begin{eqnarray*}
\sum_{d=1}^{}dp(d)= \frac{D}{N_a}, \quad Ce^{\alpha} \sum_{d=1}^{\infty}e^{\alpha}d\frac{e^{-\alpha d}}{(1 - e^{-\beta})^d} = \frac{D}{N_a}.
\end{eqnarray*}

Developped by $\frac{x}{(1 - x)^2} = \sum_{k=1}^{\infty}kx^{k}$ it follows that:

\begin{eqnarray*}
Ce^{-\alpha}\frac{\frac{e^{-\alpha}}{1 - e^{-\beta}}}{\left(1 - \frac{e^{-\alpha}}{1 - e^{-\beta}}\right)^2} = \frac{D}{N}, \quad
C\frac{1 - e^{-\beta}}{(1 - e^{-\beta} - e^{-\alpha})^2} = \frac{D}{N_a}.
\end{eqnarray*}

Concerning the average wealth we have:

\begin{eqnarray*}
\sum_{m=0}^{\infty} m p(m) = \frac{M}{N_a}, \quad C e^{\beta}\sum_{m=0}^{\infty} m \frac{e^{-\beta (m+1)}}{(1 - e^{-\alpha})^{m+1}} = \frac{M}{N_a}.
\end{eqnarray*}

That requires:

\begin{eqnarray*}
0< \frac{e^{-\beta}}{1 - e^{-\alpha}}<1, \quad C\frac{1}{1 - e^{-\alpha}}\sum_{m=0}^{\infty} m \frac{e^{-\beta m}}{(1 - e^{-\alpha})^{m}} = \frac{M}{N_a}, \quad \\
C\frac{1}{1 - e^{-\alpha}} \frac{\frac{e^{-\beta}}{1 - e^{-\alpha}}}{\left(1 - \frac{e^{-\beta}}{1 - e^{-\alpha}}\right)^2} = \frac{M}{N_a}, \quad C\frac{e^{-\beta}}{(1 - e^{-\beta} - e^{-\alpha})^2} = \frac{M}{N_a}.
\end{eqnarray*}

By $z = C$, $x = e^{-\alpha}$ e $y = e^{-\beta}$ the problem can be approached by the following system of equations:

\begin{eqnarray*}
\left\{ 
    \begin{array}{l}
        \frac{z}{1-x-y} = 1\\
        \frac{(1-y)z}{(1-x-y)^2} = \frac{D}{N_a}\\
        \frac{yz}{(1-x-y)^2} = \frac{M}{N_a}\\
    \end{array} 
\right.
\quad
\left\{ 
    \begin{array}{l}
        x = \frac{D-N_a}{D+M}\\
        y = \frac{M}{D+M}\\
        z = \frac{N_a}{D + M}.\\
    \end{array}
\right.
\end{eqnarray*}

It follows: 

\begin{eqnarray*}
C=\frac{N_a}{D+M}, 	\quad \alpha = -ln\left(\frac{D - N_a}{D + M}\right), \quad  \beta = -ln\left(\frac{M}{D+M}\right).
\end{eqnarray*}

One has to observe that:

\begin{eqnarray*}
e^{-\alpha} + e^{-\beta} = \frac{D-M}{D+M} + \frac{M}{D+M} = 1 - \frac{N_a}{D+M}, 
\end{eqnarray*}

is lower than 1.
Therefore, the convergence condition is always satisfied regardless of the specific values that fulfil the condition $N_a < D << M$.


\bibliographystyle{elsarticle-harv}






\end{document}